\documentclass[preprint,showpacs,preprintnumbers,amsmath,amssymb]{revtex4}%
\usepackage{graphicx}
\usepackage{dcolumn}
\usepackage{bm}
\usepackage{amsmath}
\usepackage{amsfonts}
\usepackage{amssymb}%
\begin{document}
\preprint{APS/123-QED}
\title{Simulated Dynamical Weakening and Abelian Avalanches in Mean-Field Driven
Threshold Models}
\author{Eric F. Preston}
\email{efp@newton.indstate.edu}
\affiliation{Department of Physics, Indiana State University, Terre Haute, IN 47809}
\author{Jorge S. S\'{a} Martins}
\affiliation{Instituto de F\'{\i}sica, Universidade Federal Fluminense, Campus da Praia
Vermelha, Boa Viagem, Niter\'oi, RJ, Brazil, 24210-340}
\author{John Rundle}
\affiliation{Center for Computational Science and Engineering, One Shields Avenue,
University of California, Davis, CA 95616}

\begin{abstract}
Mean-field coupled lattice maps are used to approximate the physics of driven
threshold systems with long range interactions. However, they are incapable of
modeling specific features of the dynamic instability responsible for
generating avalanches. Here we present a method of simulating specific
frictional weakening effects in a mean field slider block model. This provides
a means of exploring dynamical effects previously inaccessible to discrete
time simulations. This formulation also results in Abelian avalanches, where
rupture propagation is independent of the failure sequence. The resulting
event size distribution is shown to be generated by the boundary crossings of
a stochastic process. This is applied to typical models to explain some
commonly observed features.

\end{abstract}

\date{\today}
\maketitle
\tableofcontents

\section{Introduction}

In the study of earthquakes, computational slider block models
\cite{Burridge67,Rundle77,Nakanishi90,Ferguson97} are often used to
investigate the origin of magnitude-frequency scaling, a ubiquitous feature of
global seismicity. Slider block models are most commonly implemented as a
coupled lattice map, a system of continuous variables interacting in discrete
time. Similar driven threshold models are used to describe a wide range of
systems, such as pinned charge density waves \cite{Fisher85}, flux lattices in
type II superconductors \cite{Giamarchi95}, and creeping contact lines
\cite{Ertas94}. All these systems exhibit instabilities which form complex
spatiotemporal patterns, usually with a regime power law events.

Driven threshold systems with long-range interactions are common in nature,
but difficult to investigate computationally. However, the limiting case of
mean-field models are beginning to yield to analytical understanding
\cite{Dahmen98,Klein99,Preston00,Ding93}. Mean field model simulations exhibit
complex event histories and regimes of behavior, including a power law
magnitude-frequency relation. Despite their simplicity, mean-field models
remain sensitive to the choice of update rules and details of implementation.
This is especially evident in models that impose a particular form of
frictional weakening, where different modes of behavior appear as the strength
of the weakening is varied \cite{Dahmen98}.

Frictional weakening refers to the drop in cohesive force with velocity or
slip, generating the dynamical instability which produces an avalanche. It is
an essential component in the dynamics of rupture, but requires continuous
time dynamics to simulate. This is prohibitively expensive, severely limiting
the scale of the models one can investigate. Current coupled lattice map
models necessarily assume the final state does not depend on the weakening
law, in contradiction to the earliest findings in computational seismology
\cite{Burridge67}.

Some coupled lattice map models attempt to simulate frictional weakening with
modified update rules, generating new species of models with their own
behavioral quirks \cite{Ben-Zion93}. This form of weakening is rigid in design
and qualitatively different from the real phenomenon. In addition, this method
compounds analytical difficulties involving multiple failures of individual
blocks in a single step \cite{Klein99}.

Here we present a method of introducing realistic weakening effects in
discrete time simulation.\ We have developed techniques of simulation and
analysis which encompass arbitrary weakening laws under identical rules of
evolution. These techniques should be applicable to a wide range of driven
threshold systems, where more precise behavior of the instability may be
modeled.\ This unifies the analysis of previously incompatible models, and
provides more freedom in numerical simulation. This technique also results in
Abelian rupture propagation, where the size of a simulated earthquake is
uniquely determined from initial conditions, leading to a rigorous and
implementation independent analysis. We then present this analysis for mean
field slider block models, and discover a theoretical description of observed
finite-size effects usually interpreted as a nucleation-type phenomenon.

We first review the basic details of a mean-field slider block model to
demonstrate the origin of algorithmic dependence. We then present the
technique of `forced weakening', which uses a stress excess function, to
account for the macroscopic effects of an arbitrary microscopic weakening law.
This technique is used to understand the relationships between current models,
and provides a means to explore dynamical effects previously inaccessible to
discrete time simulation.\ Finally, we use this technique in model analysis to
understand the behavior of mean-field models with new accuracy and generality.

\section{The Near Mean Field Model}

The general slider-block model represents stick-slip motion along a fault
plane with $N\gg1$ discrete coordinates (or `sites') coupled by springs. Each
site is assigned a slip deficit $u_{i}$ which measures the distance from
global elastic equilibrium. The sites are pinned in place by frictional
forces, and are subject to a restoring force (which is traditionally called
`stress') proportional to their slip deficit. All sites are subject to an
external driving force which uniformly increases the slip deficits. Internal
disorder gives rise to an additional component of stress through site
interactions. The stress $S_{i}$ at a site $i$ is related to the slip deficits
through a linear constitutive relation%

\begin{equation}
S_{i}=-K_{L}u_{i}-\sum_{j}K_{ij}\left(  u_{i}-u_{j}\right)
\end{equation}
where $K_{L}$ and $K_{ij}$ are spring constants. If we impose uniform
(mean-field) interactions between all the elements, $K_{ij}=K_{C}/N$, the
above relation becomes
\begin{equation}
S_{i}=-K_{L}u_{i}-K_{C}(u_{i}-\left\langle u\right\rangle )
\label{cont eq with units}%
\end{equation}
where $\ \left\langle u\right\rangle =N^{-1}\sum_{i}u_{i}$ will denote an
average over all $N$ sites in the model. We obtain a unitless expression by
dividing by $K_{C}a$, where $a$ is a characteristic microscopic length (the
equilibrium distance between sites). Defining the unitless slip deficit
$\phi=u/a$, stress $\sigma=S/(K_{c}a)$, and spring constant ratio $K_{R}%
=K_{L}/K_{C}$, (\ref{cont eq with units}) simplifies to
\begin{equation}
\sigma_{i}=\left\langle \phi\right\rangle -(K_{R}+1)\phi_{i}. \label{const eq}%
\end{equation}
For finite $N$ we will refer to this as the near mean-field model. Note that
it is easy to invert (\ref{const eq}) for the slip deficits in terms of
stresses,
\begin{equation}
\phi_{i}=\frac{-\sigma_{i}}{K_{R}+1}-\frac{\left\langle \sigma\right\rangle
}{K_{R}(K_{R}+1)}%
\end{equation}
so the configuration is uniquely determined by the parameter $K_{R}$ and
either the slip deficits or stresses alone.

The model is slowly driven away from equilibrium by uniformly increasing all
slip deficits. Eventually the stress at one site will surpass the maximum
local frictional force and `fail', sliding toward its equilibrium point. The
motion of a failed site will change the mean slip deficit $\left\langle
\phi\right\rangle $, and produce a change in stress at other sites. If this
change brings other sites to their threshold, they will also fail, producing
an avalanche interpreted as a single event.

This description assumes that when a site begins to slip, the frictional force
weakens, producing a transient dynamic instability. In discrete time we cannot
model the dynamic slip or velocity of the site, but instead assign a residual
stress $\sigma^{R}$ at which the motion arrests. This $\sigma^{R}$ is chosen
from a probability distribution independently for each failed site. Since
slips occur instantaneously, we lose the interplay between a continuously
evolving stress field and frictional force at a site. The behavior of
dynamical models is known to strongly depend on the form of frictional
weakening \cite{Burridge67}, a feature entirely absent from coupled lattice maps.

Since the near mean field model is not dynamical we are only interested in
large-scale features of its behavior that are independent of microscopic
dynamics. Thus we are free to choose the simplest update rules that are
consistent with fracture processes. In practice, we assume that a single site
$j$ reaches its stress threshold first. Since the physics will depend only on
changes in stress, we may impose a uniform failure threshold $\sigma^{F}$ by
absorbing threshold variations into the residual stress distribution. \ The
slip displacement $\Delta_{j}=\phi_{j}^{(f)}-\phi_{j}^{(i)}$ is related to the
change in stress $\Delta\sigma_{j}=\sigma_{j}^{R}-\sigma^{F}$ by $\Delta
_{j}=-\Delta\sigma_{j}/(K_{R}+1-N^{-1})$. The motion of the site will change
the mean slip deficit $\left\langle \phi\right\rangle $ by $\Delta_{j}/N$. We
may interpret this as a transfer of stress from failing sites to pinned sites.

\subsection{Model Behavior}

\begin{figure}
[ptb]
\begin{center}
\includegraphics[
height=4.5637in,
width=5.9153in
]%
{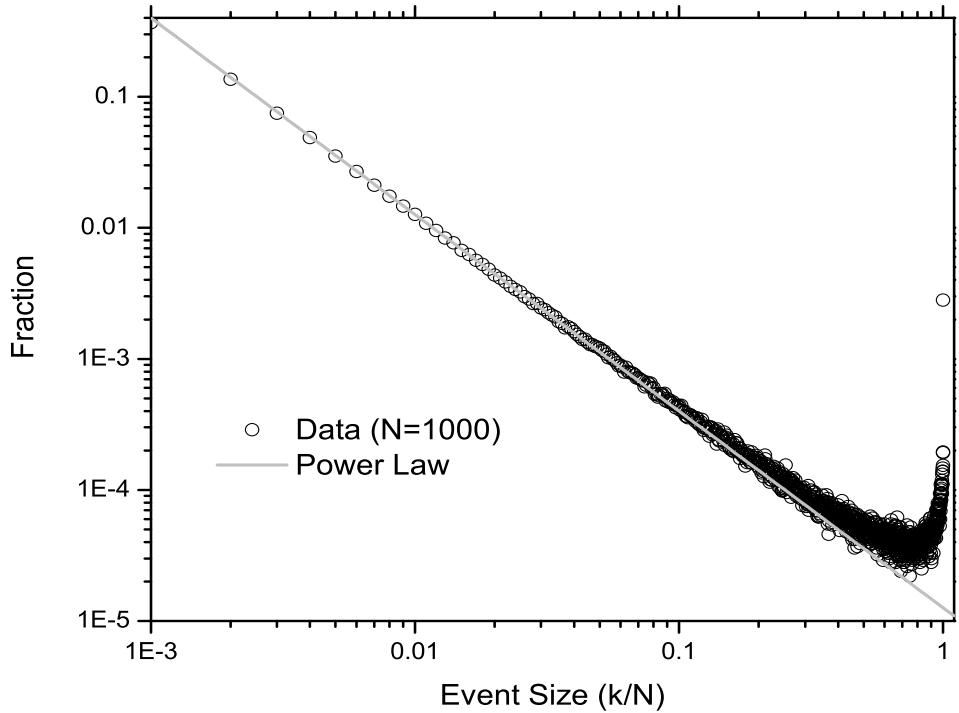}%
\caption{Distribution of event sizes for near mean field model simulation with
$K_{R}=0$. Note the overabundance of large events.}%
\label{Fig: Power Law}%
\end{center}
\end{figure}

The behavior of this model is observed through numerical simulation and
depends on the constant $K_{R}$, the distribution of residual stresses
$\sigma^{R}$, the initial conditions, and possibly on special weakening rules.
For large $K_{R}$ the coupling becomes unimportant and the system acts as $N$
independent stick-slip blocks. With small $K_{R}$ and generic (randomized)
initial conditions the model exhibits a power law in event sizes with the
mean-field exponent of 3/2 (Fig. \ref{Fig: Power Law}). It is this apparently
critical behavior which has drawn attention to this model as an analogue to
earthquakes and other largely scale-invariant phenomena.

\begin{figure}
[ptbptb]
\begin{center}
\includegraphics[
height=4.562in,
width=5.9601in
]%
{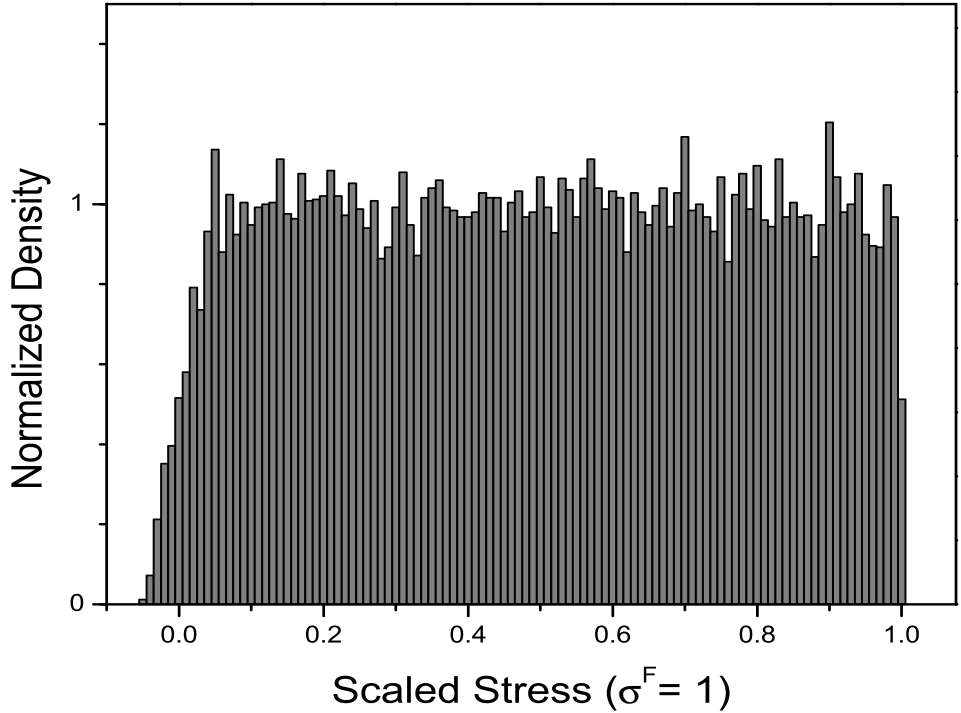}%
\caption{A histogram of stress values of sites in a typical near mean field
simulation with uniform random residual stress between -0.05 and 0.05. The
distribution has the statistical properties of a set of uniformly distributed
random variables between 0.05 and 1.}%
\label{Fig: Stress Density}%
\end{center}
\end{figure}

In critical models the events are uncorrelated in time or magnitude. Events in
any magnitude range occur as a Poisson point process in time with a rate
appropriate to their relative abundance \cite{Preston01}. The stresses in the
model at any time appear uniformly distributed between the upper bound of the
residual stress and the failure threshold. An example stress distribution is
shown as a histogram in Fig. \ref{Fig: Stress Density}.%

\begin{figure}
[ptb]
\begin{center}
\includegraphics[
height=4.562in,
width=5.9601in
]%
{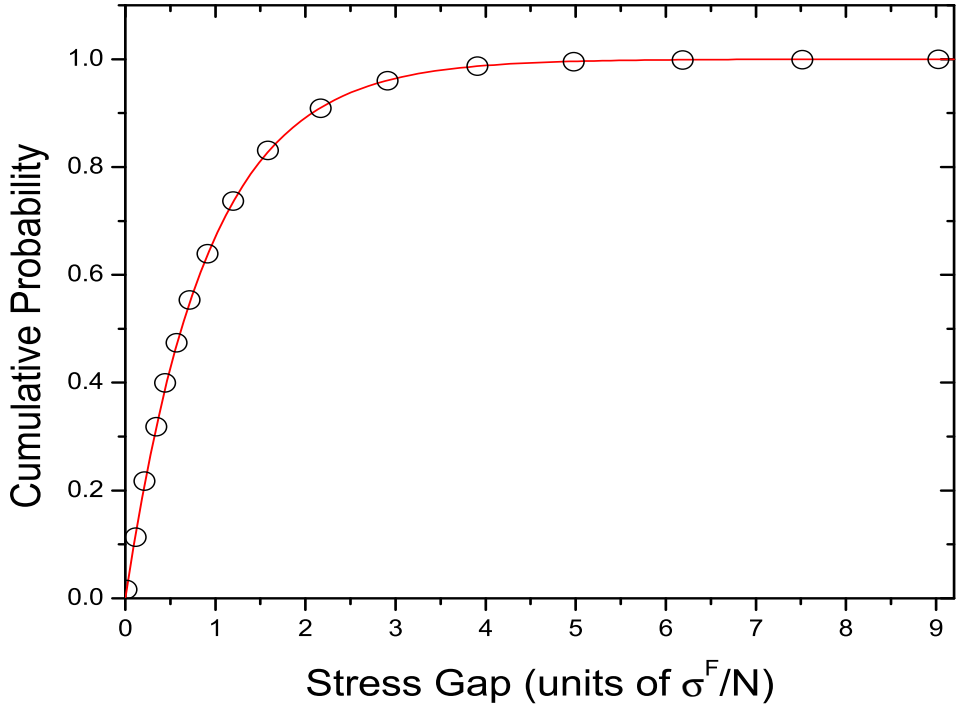}%
\caption{Comparison of simulated stress gap distribution with expectation from
assuming independent uniformly distributed values. Plotted is the cumulative
distribution of stress gaps (distance between nearest stress values) vs.
cumulative exponential distribution.}%
\label{Fig: Gap Distribution}%
\end{center}
\end{figure}

If we assume the stress value of each site is independently sampled from such
a uniform distribution, and order the results by increasing stress, we expect
the stress gaps between nearest values to have an exponential distribution.
Comparison of this theoretical distribution with an actual simulation (Fig.
\ref{Fig: Gap Distribution}) show the assumption remarkably valid for large
$N$. As long as there is a very small (Order $N^{-1}$) randomized residual
stress, these stress statistics will be persistent in time and independent for
each event. With no randomizing ingredient, limit cycles with time correlated
distributions are possible.%

Many ideas concerning the overabundance of large events in Fig.
\ref{Fig: Stress Density} have been put forward. This has often been
considered an example of characteristic behavior, that is, repetition of an
event whose size is determined by the local geometry. This can be considered a
type of nucleation phenomena, a runaway Griffith rupture, where events larger
than a critical size can not arrest.

This characteristic behavior is emphasized in dynamic weakening models
\cite{Ben-Zion93}, where the increased propensity for a failed site to slip
further (due to a weakened pinning force) is simulated by imposing a lower
threshold stress $\sigma^{D}<\sigma^{F}$ for the duration of a single event.
After a site fails it will receive stress transfer from subsequent failures,
and thus may reach this lower threshold and fail again. Re-failing sites will
contribute more to the external transfer and enhance the likelihood of
continued rupture growth.

This form of weakening manifests itself after some failed sites have their
stress brought back up to the dynamical threshold. This will only occur when
the rupture reaches a certain size.\ After the onset of this dynamical
weakening, the additional stress transfer typically results in a runaway event
encompassing every block in the system. Thus, this form of weakening tends to
produce characteristic events which always occur once the minimum rupture size
is reached. To understand how the stress distribution and weakening rules
determine the event size distribution and model behavior, we must examine the
details of stress transfer between sites.

\subsection{Stress Transfer}

The above describes a series type dislocation where sites fail in sequence and
the stress transfer occurs to other sites all at once. This makes it likely
that any failing site (other than the single initiator) will have a stress
slightly \emph{above} the threshold, which subtly provides an order-of-failure
dependence to the stress transfer. As a consequence, the exact stress transfer
in simulation will depend on obscure factors like the order of iteration over
sites. To eliminate this we must examine the stress transfer in more detail.

Suppose that in the course of an event there have been $k$ block failures. Let
$\{k\}$ represent the set of indices of failed sites. Call $\kappa=k/N$ the
fraction of failed sites. Then from (\ref{const eq}) the change in stress for
any stable site $i$ is%

\begin{align}
\Delta\sigma_{i\notin\{k\}}  &  =\frac{1}{N}\sum_{j\in\{k\}}\Delta_{j}%
=\frac{\delta}{N}\sum_{j\in\{k\}}(\sigma_{j}^{f}-\sigma_{j}^{R})\nonumber\\
&  =\delta\kappa\left(  \left\langle \sigma^{f}\right\rangle _{k}-\left\langle
\sigma^{R}\right\rangle _{k}\right)  =\delta\kappa\left\langle -\Delta
\sigma^{f}\right\rangle _{k} \label{Stress Transfer A}%
\end{align}
where $\delta=(K_{R}+1-N^{-1})^{-1}$, and $\left\langle \cdot\right\rangle
_{k}$ is an average applied over failed sites. We call this the \emph{external
stress transfer} to signify that it applies to sites that are not part of the
rupture. The quantity $\sigma_{j}^{f}$ is the stress of site $j$ \emph{at
failure}, which may be greater than $\sigma^{F}$. Note that the slip
displacement $\Delta_{j}$ is only dependent on the stress drop at failure
because pinning occurs immediately, and subsequent stress changes will not
cause this site to slip further. Since $\sigma^{f}$ is typically very near
$\sigma^{F}$ and the $\sigma_{j}^{R}$ are identically distributed random
variables, the term in parenthesis is on average independent of $k$. Thus the
external transfer grows linearly with the fraction of failed sites.

To examine the effects of stress overshoot, suppose a site $j$ slips on the
initial step. The external transfer is
\begin{equation}
\sigma_{i\neq j}=\sigma_{i}^{0}+\frac{\delta\left(  \sigma_{j}^{0}-\sigma
_{j}^{R}\right)  }{N} \label{stress change from slip}%
\end{equation}
where for emphasis the superscript $0$ denotes the \emph{initial} stress at a
site before any failures have occurred. Of course, for the rupture initiator,
$\sigma_{j}^{0}=\sigma^{f}=\sigma^{F}$. Now suppose this transfer causes sites
$m$ and $n$ to fail. At this time $\sigma_{m,n}=\sigma_{m,n}^{0}+\delta\left(
\sigma_{j}-\sigma_{j}^{R}\right)  /N$, so after $m$ and $n$ fail the external
transfer becomes
\begin{equation}
\sigma_{i\neq\{j,m,n\}}=\sigma_{i}^{0}+\frac{\delta\left(  \sigma_{j}%
^{0}-\sigma_{j}^{R}\right)  }{N}+\frac{\delta\left(  \sigma_{m}^{0}-\sigma
_{m}^{R}\right)  }{N}+\frac{\delta\left(  \sigma_{n}^{0}-\sigma_{n}%
^{R}\right)  }{N}+\frac{2\delta^{2}\left(  \sigma_{j}^{0}-\sigma_{j}%
^{R}\right)  }{N^{2}}. \label{naive transfer}%
\end{equation}
Note the factor of $2$ in the last term above comes from the fact that two
sites ($m,n$) failed in the previous iteration. This simple stress transfer
rule introduces dependence on which blocks fail during which iteration step.
We would prefer the time evolution to be a transparent function of the initial
stresses only. To satisfy this condition, we must make the stress transfer
Abelian, that is, independent of the order of failure. This is similar in
concept to the Abelian sand pile model \cite{Dhar90,Dhar99}. The key to making
stress transfer Abelian is to use the additional degree of freedom offered
with the inclusion of simulated weakening.

\subsection{Forced Weakening}

\begin{figure}
[ptb]
\begin{center}
\includegraphics[
height=4.5645in,
width=5.9003in
]%
{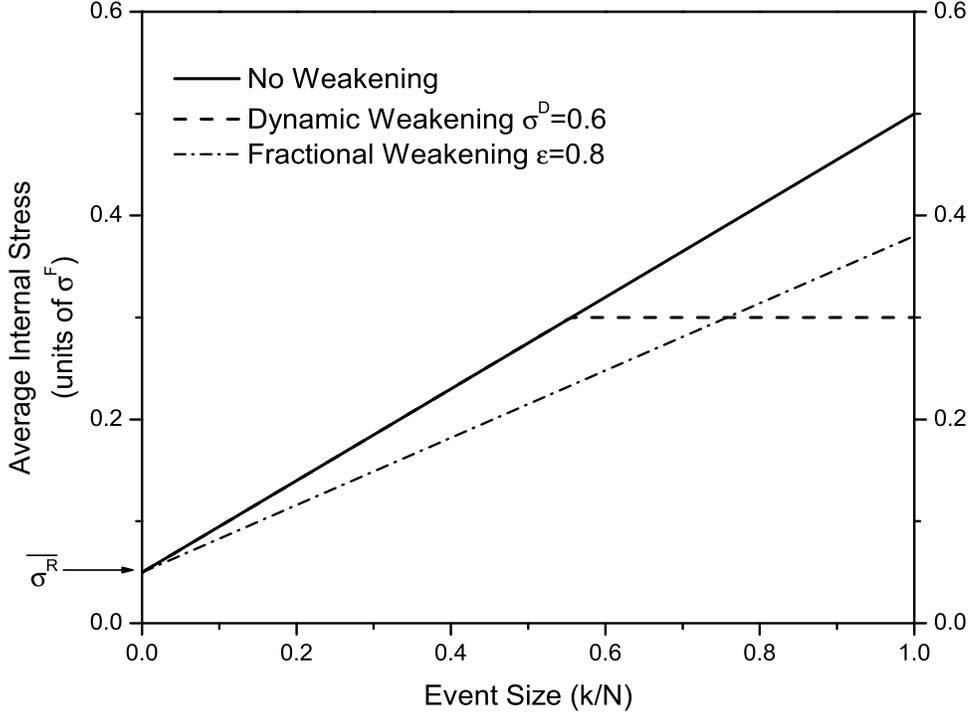}%
\caption{Schematic of average internal stress (average stress of sites that
have failed) as a function of rupture size for several methods of simulated
weakening.}%
\label{Fig: aisfuncs}%
\end{center}
\end{figure}

One way to visualize the effects of weakening is to examine the average stress
of sites that have failed as a function of rupture size. After failure, a site
has average stress $\left\langle \sigma^{R}\right\rangle $. Subsequent
failures will transfer some stress to this now pinned site. Without weakening,
the average stress of failed sites will grow with rupture size $\kappa$ as%
\begin{equation}
\overline{\sigma_{int}}(\kappa)=\frac{\delta}{2}(\kappa-N^{-1})\left\langle
\Delta\sigma\right\rangle _{k}.\label{no weak ais}%
\end{equation}
With dynamic weakening, all failed sites with stress $\geq\sigma^{D}$ will
fail again, putting a ceiling on the average internal stress, as illustrated
in Fig. \ref{Fig: aisfuncs}.%

We might imagine a different approach to weakening where failed sites shed a
certain fraction of the stress they receive after failure by continuing to
slide. This would not be convenient to simulate, but clearly possible. This
would produce an average internal stress function the same as one for larger
$K_{R}$, however now the stress difference is not dissipated, but transferred
to external sites. The main point of this would be to generate a weakening
effect present for all event sizes. \ Implementing this `fractional weakening'
would involve re-computing the slip of all failed sites with each new failure.
Instead, we might seek to formulate the model so the average internal stress
function is given as a physical parameter and the requisite slips and stress
transfer computed as a result. In essence, in place of deriving a weakening
function from dynamical equations involving slip or velocity dependent
friction, we can impose the effects of the weakening as they appear to stable
sites. We call this approach \emph{forced weakening}.

To arrive at a forced weakening formulation we first observe the change in
stress of a site $j$ as it depends on $k$, the number of failed sites. Let
$\Delta_{j}^{k}$ denote the slip displacement of site $j$ after $k$ failures.
If it is nonzero it includes all block motion, including initial failure,
additional failures from weakening, or continuous sliding. Consider the system
of equations for the stress changes of the failed sites $j\in\{k\}$%
\begin{align}
\Delta\sigma_{j\in\{k\}}^{k}  &  =\sigma_{j}^{k}-\sigma_{j}^{0}=-\delta
\Delta_{j}^{k}+\frac{1}{N}\sum_{i\neq j\in\{k\}}\Delta_{i}^{k}\nonumber\\
&  =-(K_{R}+1)\Delta_{j}^{k}+\frac{1}{N}\sum_{i\in\{k\}}\Delta_{i}^{k}
\label{Local Drops}%
\end{align}
where $\sigma^{k}$ denotes the stress after $k$ failures (and $\sigma^{0}$ is
the initial value). The last line demonstrates the simple linear form of the
relationship between stress drops and slip displacements. This may be obtained
via a matrix with diagonal elements $N^{-1}-(K_{R}+1)$ and off-diagonal
elements $N^{-1}$. This matrix is easily inverted to obtain the slips
$\Delta_{j}^{k}$ in terms of the current stress drops $\Delta\sigma_{j}^{k}$%
\begin{align}
\Delta_{j}^{k}  &  =\frac{-(K_{R}+1-k/N)\Delta\sigma_{j}^{k}-N^{-1}\sum
_{i\in\{k\}}\Delta\sigma_{i}^{k}}{(K_{R}+1)\left(  K_{R}+1-k/N\right)
}\nonumber\\
&  =\frac{-\Delta\sigma_{j}^{k}}{K_{R}+1}-\frac{(k/N)\ \left\langle
\Delta\sigma^{k}\right\rangle _{k}}{(K_{R}+1)\left(  K_{R}+1-k/N\right)  }
\label{Stress Transfer 2}%
\end{align}
This expression provides the slips that are necessary to generate a set of
stress drops.

The slips are directly related to the external transfer, as in
(\ref{Stress Transfer A}). Summing over them yields a new expression for the
external transfer in terms of the stress drops.
\begin{align}
\Delta\sigma_{i\notin\{k\}}  &  =\frac{1}{N}\sum_{j\in\{k\}}\Delta_{j}%
=\frac{k}{N}\left\langle -\Delta\sigma^{k}\right\rangle _{k}\nonumber\\
&  \times\left[  \frac{1}{K_{R}+1}+\frac{k/N}{(K_{R}+1)(K_{R}+1-k/N)}\right]
\nonumber\\
\Delta\sigma_{i\notin\{k\}}  &  =\frac{\kappa}{(K_{R}+1-\kappa)}\left\langle
-\Delta\sigma^{k}\right\rangle _{k} \label{final stress transfer}%
\end{align}
This is identical to (\ref{Stress Transfer A}) when $k=1$. However, the new
effective transfer coefficient $\delta(\kappa)=(K_{R}+1-\kappa)^{-1}$ grows
with the rupture size. Making up for this is the fact that the stress drops
are no longer computed only at failure, but account for stress changes
occurring as the rupture progresses. As failed sites pin and acquire
additional stress transfer, the average change in stress will decrease.
Observe that to calculate the external transfer, we need not specify
individual stress drops or slips, but require only the \emph{average dynamic
stress drop} of all failed sites. This evolving average stress drop is defined
as
\begin{align}
\left\langle -\Delta\sigma^{k}\right\rangle _{\kappa}  &  =\frac{1}{k}%
\sum_{j\in\{k\}}(\sigma_{j}^{0}-\sigma_{j}^{k})\nonumber\\
&  =\left\langle \sigma^{0}\right\rangle _{\kappa}-\overline{\sigma_{int}%
}(\kappa) \label{final stress drop}%
\end{align}
where we have defined the average internal stress (AIS) function
$\overline{\sigma_{int}}(\kappa)=\left\langle \sigma^{k}\right\rangle _{k}$,
which characterizes the frictional weakening. The stress transfer now depends
only on the average initial stress of failed sites (which decreases with
rupture size) and the AIS function. Both may be defined in terms of rupture
size independent of a particular update algorithm. Using this formulation,
avalanches are Abelian.

In numerical simulation, we must eventually assign an actual slip and/or
residual stress to each failed site. Care must be taken to make results agree
with the analogous forward simulation. For example, given a linear AIS
function $f(\kappa)=\alpha\kappa+\beta$, we could assign the $k^{th}$ failed
site a random residual stress (with mean $\beta$) plus $2\alpha(k-1)/N$. When
using AIS functions with no forward equivalent, the method of assigning final
stresses must be stated explicitly.

The forced weakening method has two main advantages. Practically, it allows
the simulation of models with arbitrary weakening characteristics, most of
which would not be obtainable with modified CA rules. Formally, the model is
Abelian, so that the event size is a unique function of the initial stress
configuration. Using this fact we can seek an expression which will determine
the event size given adequate information of the initial stresses.

\section{Model\ Analysis}

With an Abelian stress transfer, the final event size depends only on the
initial stress distribution. We could therefore compute the final event size
if we concretely characterize the stress distribution. For model solutions
we\ find it convenient to mostly work with the \emph{stress deficits} defined
as
\begin{equation}
\Sigma_{i}(t)=\sigma^{F}-\sigma_{i}(t)
\end{equation}
representing the distance a site $i$ is from failure. A rupture originates at
sites where $\Sigma=0$ and advances through sites with progressively larger
values of $\Sigma$. Consider the cumulative distribution $P_{\Sigma}(\chi)$
defined as the fraction of sites with stress deficit $\Sigma\leq\sigma^{F}%
\chi$ (scaled so $\chi$ varies from zero to one). Typically, one would choose
the microscopic length $a$ such that $K_{C}a=\sigma^{F}$ so that $\sigma
^{F}=1$. For a specific stress configuration $P_{\Sigma}(\chi)$ will be a
piece-wise continuous function with $N$ steps.

\subsection{Solution for Rupture Size}

To express the stress transfer as function of rupture size, the discrete
initial stresses $\sigma_{j}$ must be expressed in terms of the stress
distribution. Beginning with the forced weakening form of the external
transfer $\tau(\kappa)$,
\begin{equation}
\tau(\kappa)=\frac{\kappa}{K_{R}+1-\kappa}\left\langle -\Delta\sigma
^{k}\right\rangle _{\kappa}. \label{FW transfer func}%
\end{equation}
where $\left\langle -\Delta\sigma^{k}\right\rangle _{\kappa}=k^{-1}\sum
_{j\in\{k\}}(\sigma_{j}^{0}-\sigma_{j}^{k})$, we notice that the $k^{th}$ site
to fail in the rupture has a stress deficit of $P_{\Sigma}^{-1}(\frac{k-1}%
{N})$. The factor $k-1$ reflects the fact that the first site to fail has a
stress deficit of zero, so $P_{\Sigma}(0)=N^{-1}$, and $P_{\Sigma}^{-1}%
(\eta<N^{-1})$ is undefined. This relationship is illustrated graphically in
Fig. \ref{Fig: cumdist}. Using this fact we may write the average stress drop
in (\ref{FW transfer func}) as
\begin{align}
-\left\langle \Delta\sigma^{k}\right\rangle _{k}  &  =\frac{1}{k}\sum
_{j\in\{k\}}(\sigma_{j}^{0}-\sigma_{j}^{k})=\sigma^{F}-\frac{1}{k}\sum
_{j\in\{k\}}\left(  \sigma^{F}-\sigma_{j}^{0}+\sigma_{j}^{k}\right)
\nonumber\\
&  =\sigma^{F}-\frac{1}{\kappa}\sum_{j\in\{k\}}\frac{1}{N}\left(  \Sigma
_{j}^{i}+\sigma_{j}^{k}\right) \nonumber\\
&  =\sigma^{F}-\frac{\sigma^{F}}{\kappa}\int_{\varepsilon}^{\kappa}P_{\Sigma
}^{-1}(\eta-\varepsilon)d\eta-\overline{\sigma_{int}}(\kappa)
\label{Cont stress drop}%
\end{align}
where we have identified the average internal stress function $\left\langle
\sigma^{k}\right\rangle _{k}=\overline{\sigma_{int}}(\kappa)$ and introduced
the notation $\varepsilon=N^{-1}$. Note that we took no limit in writing the
integral, but recognized the equivalent of the sum and an integral over a step function.

Using this form of the average stress drop we may write the external transfer
in terms of the stress distribution as
\begin{equation}
\tau(\kappa)=\frac{1}{K_{R}+1-\kappa}\left[  \kappa\sigma^{F}-\kappa
\overline{\sigma_{int}}(\kappa)-\sigma^{F}\int_{\varepsilon}^{\kappa}%
P_{\Sigma}^{-1}(\eta-\varepsilon)d\eta\right]  . \label{Cont Stress Transfer}%
\end{equation}

Using (\ref{Cont Stress Transfer}) and the statistical properties of
$P_{\Sigma}^{-1}(\eta)$, we should be able to calculate the size of the next
event. At any point during a sequence of failures, a rupture will arrest if
every (initial) stress deficit in the unfailed region is greater than the
current external transfer. Referring to Figure \ref{Fig: cumdist}, we see that
the stress deficit of the next site to fail is $P_{\Sigma}^{-1}(\kappa)$.
Using this we define a \emph{stress excess} $v$ which will provide a criteria
for the progress of an event in terms of the stress distribution:%
\begin{figure}
[ptb]
\begin{center}
\includegraphics[
height=4.0074in,
width=5.9975in
]%
{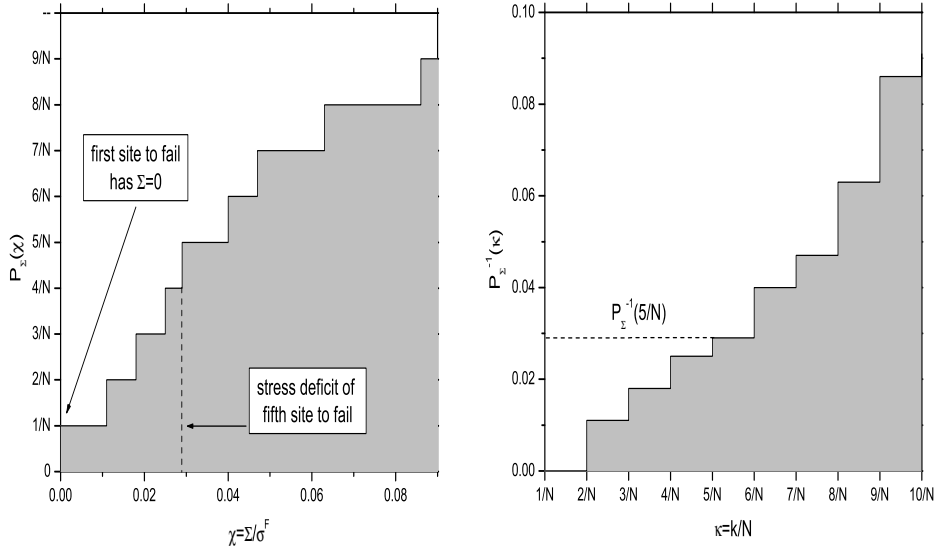}%
\caption{Relationship between inverse cumulative stress deficit function
$P_{\Sigma}$ and event size $\kappa=k/N$. A rupture must initiate at a site
where $\Sigma=0$, so the stress deficit of the $k^{th}$ site to fail is given
by $P_{\Sigma}(\kappa-N^{-1})$.}%
\label{Fig: cumdist}%
\end{center}
\end{figure}
\begin{equation}
v(\kappa)=\tau(\kappa)-\sigma^{F}P_{\Sigma}^{-1}(\kappa
).\label{stress excess function}%
\end{equation}
\ When $v$ is positive, a rupture will continue to grow, arresting only
if/when $v(\kappa)=0$. Assuming a rupture has initiated, setting the stress
excess to zero yields an integral equation for the final rupture size $\kappa
$:
\begin{equation}
\frac{1}{K_{R}+1-\kappa}\left[  \kappa\sigma^{F}-\kappa\overline{\sigma_{int}%
}(\kappa)-\sigma^{F}\int_{\varepsilon}^{\kappa}P_{\Sigma}^{-1}(\eta
-\varepsilon)d\eta\right]  -\sigma^{F}P_{\Sigma}^{-1}(\kappa
)=0\label{rupture size}%
\end{equation}
To solve this equation it would be convenient to consider $\kappa$ a truly
continuous variable. The stress distribution could vary continuously if we let
$N\rightarrow\infty$, or by averaging the stress deficits over a short time
interval. The latter works if we imagine the system driving and stress
transfers to be continuous, so all intermediate values of stress are passed
through during updates. This form of temporal course-graining has been central
to previous analysis of slider-block models \cite{Klein99}.

\subsection{Example Solutions}

It is instructive to examine the stress excess function and the event size
solutions when we can perform the integral over the step function
analytically; that is, when the steps are of equal size. In this case the
stress deficit values are $\Sigma_{1}=0,$ $\Sigma_{2}=1/N$, $\Sigma_{3}=2/N$,
etc. This will also represent the case where $N\rightarrow\infty$, and the
empirical distribution of values is identical to the probability distribution
of a single variable, and the integrals are taken over a continuous variable.
However, we will not take this limit so factors of $\varepsilon=N^{-1}$ are
apparent. In this discrete formulation, the integrals become reducible sums.

As we observed, in a system exhibiting power law behavior, the stresses are
distributed as if they were selected from a uniform distribution. Fore
simplicity, we can ignore the small variation from the uniform distribution
due to random residual stresses, since this anomaly will only be encountered
for ruptures nearly the size of the system. Thus we take $p_{\Sigma}(\chi)=1$,
$P_{\Sigma}(\chi)=\kappa$, and $P_{\Sigma}^{-1}(\kappa)=\chi$.

\subsubsection{Perfect Weakening}

We will first examine the rather artificial scenario of perfect weakening,
where no additional stress beyond the residual noise may be supported by
failed sites.\ This is achieved by imposing a constant average internal stress
$\overline{\sigma_{int}}(\kappa)=0.$

The physical interpretation of this perfect weakening scenario is much like a
democratic fiber bundle model \cite{Anderson97}. Imagine a cable composed of
several parallel axial fibers under tension, with a gradually increasing load.
In this model the individual fibers have some randomly distributed strength at
which they will break. When a fiber breaks, the load it was bearing is
distributed equally among all the remaining fibers.

In the context of a slider block model, failed blocks do not re-pin during a
rupture, and continue to slide with stress changes from additional failures.
The distribution of fiber strengths corresponds to the distribution of stress
deficits in our system. The factor of $1-\kappa$ in the denominator of the
stress transfer is equivalent to the rule of sharing the load only among
unbroken fibers. This earthquake model adds the possibility of stress
dissipation through the $K_{R}$ parameter. When $K_{R}=0$ the models are equivalent.

The stress excess function for perfect weakening is shown in Fig.
\ref{Fig: SEfuncs}.\ Determining the rupture size from (\ref{rupture size})
yields the equation
\begin{equation}
\kappa^{2}=(2K_{R}-\varepsilon)\kappa\label{Uniform kappa Solution}%
\end{equation}
which has solutions for $\kappa=0$ or $\kappa=2K_{R}-\varepsilon$.\ For
$K_{R}<\varepsilon$ the stress excess is always positive and the rupture will
run away. For slightly larger values of $K_{R}$ the stress excess function
starts negative and crosses the axis a short distance from zero, as stress
dissipation kills the rupture within the first few failures. After crossing
zero the function again increases without limit. Larger values of $K_{R}$ do
not produce critical event size distributions so aren't of interest. With
perfect weakening, ruptures of significant size will never arrest.

\subsubsection{No Weakening}

\begin{figure}
[ptb]
\begin{center}
\includegraphics[
height=4.5595in,
width=5.9759in
]%
{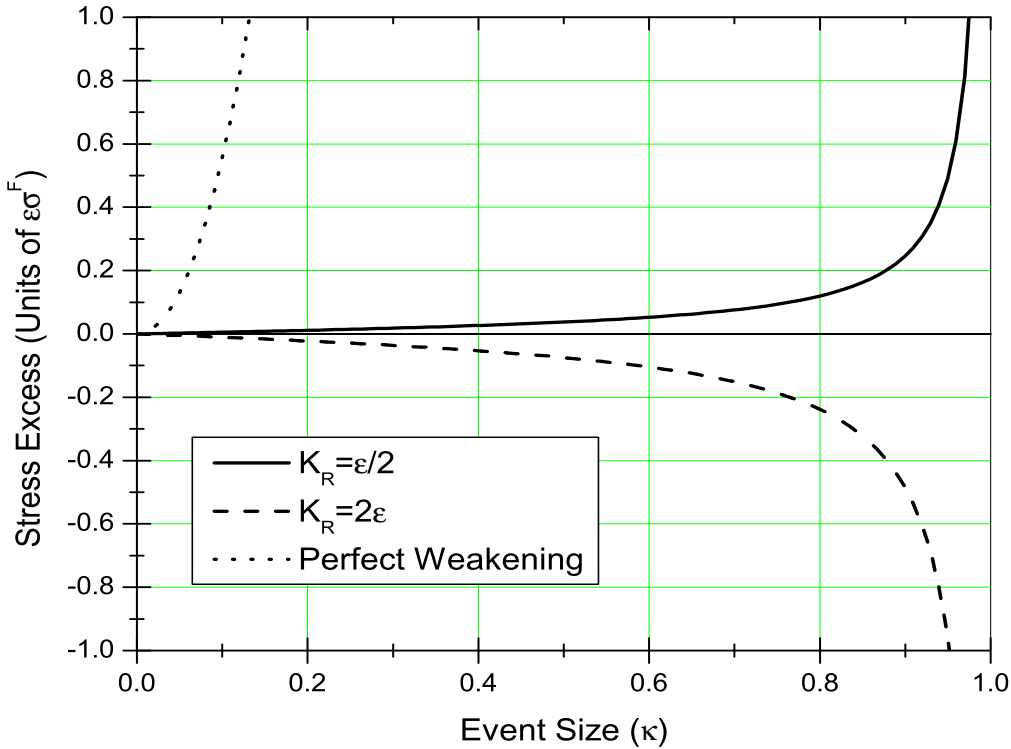}%
\caption{Ideal stress excess functions for models with perfect weakening and
no weakening.}%
\label{Fig: SEfuncs}%
\end{center}
\end{figure}

Now consider a model with a linear AIS\ function, equivalent to typical
slider-block models with no weakening:
\[
\overline{\sigma_{int}}(\kappa>0)=\delta\frac{(\kappa-\varepsilon)}{2}.
\]
The solution to the rupture size equation is now
\begin{equation}
\kappa^{2}=\kappa(2+2K_{R}-\varepsilon) \label{no weak solution}%
\end{equation}
which has solutions at $\kappa=0$ and near $\kappa=2$ for small $K_{R}$. More
important, however, is the behavior of the stress excess function below
$\kappa=1$. Again the critical value of $K_{R}$ is $\varepsilon$: for
$K_{R}<\varepsilon$, the initial stress excess is positive and remains so up
to $\kappa=1$, so again any rupture that initiates will run away (the model is
super-critical). For $K_{R}>\varepsilon$, the stress excess is always
negative, and no rupture will form (Fig. \ref{Fig: SEfuncs}).

However, when $K_{R}\rightarrow\varepsilon$ the stress excess function
vanishes for all $\kappa$. In this case, we have a critically propagating
rupture, with the exact stress necessary to tumble each site in succession.
Also, notice in Fig. \ref{Fig: SEfuncs} how closely the stress excess remains
to zero when $K_{R}$ is of order $\varepsilon$. In this case small
fluctuations in the stress distribution could generate a transient positive
stress excess, resulting in a finite rupture size.%

\subsubsection{Dynamic Weakening}

Dynamic weakening can be treated with a combination of the above two cases: as
long as the external transfer is below the dynamic stress threshold, there is
no weakening. Once external transfer crosses the dynamic threshold, the
average internal stress function becomes a constant, resulting in runaway
stress excess as in the perfect weakening scenario. The point at which this
crossover happens is easy to calculate by setting $\tau(\kappa)=\sigma
^{D}/\sigma^{F}$.\ For $W=0$ and $K_{R}=0$, this yields a crossover point at
$\kappa=\sigma^{D}/\sigma^{F}$, which is no surprise. Nonzero values of $W$ or
$K_{R}$ result in slightly larger values of the crossover point.

\subsubsection{Fractional Weakening}

The fractional weakening AIS function is identical to that of no weakening
with a modified slope. With no weakening the slope is $\delta/2$, which
depends on $K_{R}$. \ Fractional weakening can effectively balance the effect
of dissipation, moving the critical value of $K_{R}$ up. Other conclusions of
the case with no weakening remain the same.

Using these ideal solutions we can easily understand the role of the control
parameters and weakening function in terms of how a single event propagates.
Clearly, without fluctuations in the stress distribution, the model behavior
is trivial. Now we will examine solutions for finite models where the
time-averaged stress distribution is a stochastic process.

\section{Fluctuations and Stochastic Rupture Propagation}

When using a smooth stress distribution the solutions for the rupture size are
trivial, producing no complexity.\ Passing to the thermodynamic limit yields a
model with no fluctuations in the stress distribution, a true mean field
model. This is not what we observe in simulation, as actual stress
distributions are finite realizations. A model with a finite number of
elements, no matter how large, is inherently different from an infinite model.
The finite $N$ and consequent course grained empirical distribution provides
the inherent discreteness necessary to produce complex behavior, a near mean
field model.

In a short time interval $\Delta t$ where no failures occur, the stress at
each site changes an amount $\Delta\sigma=(K_{R}+1)V\Delta t$ where $V$ is a
velocity measured in units of the microscopic length $a$ per unit time. A time
average is equivalent to specifying $P_{\Sigma}(\chi)$ as the integral of a
density function $p_{\Sigma}(\chi)$, which is the average number of sites with
stress deficit $\Sigma$ between $\sigma^{F}\chi$ and $\sigma^{F}(\chi
+\Delta\chi)$, where $\Delta\chi$ is the change in scaled stress
$(\Delta\sigma/\sigma^{F})$ over the time interval. For a stress distribution
of normalized width $I$ $(0<I\leq1)$, the expected time to the next event is
defined by $\Delta\chi=I/N$, or $\Delta t=I/[NV(K_{R}+1)]$.

Based on the mean field distribution of event sizes the expected event size at
any time is $k=\sqrt{N}$. Thus, during the average time interval between
events, we expect the distribution function to average over $\sqrt{N}$
different stress values. For large $N$, this should be adequate to define a
stress distribution function as a stochastic process with well defined mean
and variance at each point. If we use this statistical description of the
stress distribution, we will obtain a statistical answer for the event size.

The near mean field model preserves the essential fluctuations characteristic
of discrete models, while also proving analytically tractable. The stress
distribution averaged over the inter-event time is a stochastic process, and
computing the event size from this will yield a probability distribution.
Since we observe the stress distribution is statistically stationary, the
distribution of event sizes for any time step is also the time-average distribution.

Again consider the case where the stresses (or stress deficits)\ appear to
have been chosen from a uniform distribution. Consider an ordering of stress
deficits $\Sigma_{1}<\Sigma_{i}<\ldots<\Sigma_{N}$ and define $\Sigma_{1}=0$
and $\Sigma_{N+1}=1$. Let $Y_{i}=\Sigma_{i+1}-\Sigma_{i}$ for $i=1\ldots N$.
Then we would expect the gaps $Y_{i}$ to be independent identically
distributed (i.i.d.) random variables with an exponential distribution of mean
$\mu_{Y}=N^{-1}$ and variance $\operatorname{Var}Y=N^{-2}$ (taking $W=0$ for
simplicity). Derivations of the event size distribution from these statistics
have been considered before \cite{Dahmen98,Ding93}, but the forced weakening
formulation provides new rigor and generality in the results. We will also
derive the finite-size correction to the mean field power law behavior for the
first time.

Notice that the partial sum over intervals is nothing but the inverse
cumulative distribution,
\begin{align}
\sum_{i=1}^{k}Y_{i}  &  =\Sigma_{k+1}-\Sigma_{1}\nonumber\\
&  =P_{\Sigma}^{-1}(\frac{k}{N})
\end{align}
yielding the stress deficit of the $k^{th}$ site to fail.

Next define the zero mean random variables
\begin{align}
X_{k/N}  &  =\frac{\sum_{i=1}^{k}(Y_{i}-\mu)}{\sigma\sqrt{N}}\nonumber\\
&  =\sqrt{N}\sum_{i=1}^{k}Y_{i}-\frac{k}{\sqrt{N}}%
\end{align}
that are partial sums of the gaps up to the $k^{th}$ site. As $N\rightarrow
\infty$ the intervals $X_{t^{\prime}}-X_{t}$ are Gaussian random variables
with mean zero and variance $t^{\prime}-t$. This is a standard method of
constructing a Wiener process from any underlying i.i.d. sequence of random
variables. Since we aren't actually taking the infinite $N$ limit, we should
consider this process equivalent to a Wiener process at scales much greater
than $N^{-1}$.

Knowing that the $X_{t}$ represents a Brownian motion in the continuum limit
we may express the inverse cumulative distribution as
\begin{align}
P_{\Sigma}^{-1}(\kappa)  &  =\sum_{i=1}^{k}Y_{i}=\left(  \frac{X_{k/N}}%
{\sqrt{N}}+\frac{k}{N}\right) \nonumber\\
&  =\frac{X_{\kappa}}{\sqrt{N}}+\kappa
\end{align}
Thus the inverse cumulative distribution may be presented as a sum of the
ideal linear term (drift) and a Wiener process with variance $N^{-1}$.\ Note
that the fluctuations scale with the system size as expected.

The stress excess function also depends on the integral of the inverse
cumulative distribution, representing the average stress deficit of failed
sites
\begin{align}
\kappa\left\langle \Sigma\right\rangle _{\kappa}  &  =\int_{0}^{\kappa
}P_{\Sigma}^{-1}(\chi)d\chi=\frac{\kappa^{2}}{2}+\frac{1}{\sqrt{N}}\int
_{0}^{\kappa}X_{\chi}d\chi\\
&  =\frac{\kappa^{2}}{2}+S_{\kappa}%
\end{align}

This integral introduces a new process $S_{\kappa}$. This is a sum of
independent Gaussian random variables with zero mean. The result is a random
variable with zero mean and a variance $\kappa^{3}/3N+\mathcal{O}(1/N^{2})$
\cite{Preston01}. This term does not bias the stochastic component, and for
large $N$ it's contribution will be negligible next to $X_{k}/\sqrt{N}$.

The stress excess function now becomes
\begin{equation}
\frac{v(\kappa)}{\sigma^{F}}=\frac{1}{K_{R}+1-\kappa}\left[  \kappa
-\kappa\frac{\overline{\sigma_{int}}(\kappa)}{\sigma^{F}}-\frac{\kappa^{2}}%
{2}-S_{\kappa}\right]  -\kappa-\frac{X_{k}}{\sqrt{N}}.
\label{Stochastic Stress Transfer}%
\end{equation}

First examine the behavior of this stochastic function near $\kappa=0$.
Consider the critical model ($K_{R}=N^{-1}$) with no weakening. In studying
(\ref{no weak solution}) we saw this function vanishes for all values of
$\kappa$. Only the noise terms, dominated by $X_{\kappa}/\sqrt{N}$, are
present. Thus, the distribution of event sizes is the distribution of
zero-crossings of a standard random walk.

\subsection{Event Size Distributions}

The zero-crossing distribution is easily derived using the fact that any
i.i.d. distribution can be used to construct the Wiener process with identical
results. It's convenient to build the walk from a discrete random variable
taking values of $\pm1$ with equal probability. Let $P_{d}(k)$ denote the
probability that the walk is $d$ steps from the origin after $k$ steps are
taken. Taking the (horizontal) step size of the walk to be $1/2N$, returns to
zero are possible at steps $k/N$ with probability
\begin{equation}
P_{0}(k)=\frac{1}{2^{2k}}\frac{(2N-2)!}{(N-1)!^{2}}%
\end{equation}
The probability of $k$ being the \emph{first} crossing of zero is $P_{0}%
(k)/k$. These probabilities have a power law distribution as seen by taking
the log and applying Stirling's approximation:
\begin{equation}
\frac{P_{0}^{k}}{k}\simeq\frac{k^{-3/2}}{N^{3/2}\sqrt{2\pi}}%
\end{equation}

The result is a power law with an exponent of $-3/2$ as observed in Fig.
\ref{Fig: Power Law}. This distribution matches the power law behavior
observed in simulation, except at the largest values.

The expected distance $d$ from the origin for an unbiased random walk in terms
of the number of steps $k$ is given by $\left\langle d_{k}\right\rangle
\simeq\sqrt{k\pi/2}$ for large $k$. With our distance per step of $1/\sqrt{N}%
$, the expected magnitude of the walk grows as $\sqrt{\pi\kappa/2}$. However,
all the stress gaps must add to one (minus the $\mathcal{O}(1/N)$ distance
$\sigma^{F}-\Sigma_{N}$), so the walk must be constrained to return to zero at
$\kappa=1$. If the stress values are considered a Poisson point process, this
is equivalent to requiring there to be exactly $N$ values between $0$ and
$\sigma^{F}$. For $N\gg1$, this constraint is unnoticeable in the statistics
of individual stress gaps, but manifests itself when all the gaps are summed.

\begin{figure}
[ptb]
\begin{center}
\includegraphics[
height=4.5637in,
width=5.9153in
]%
{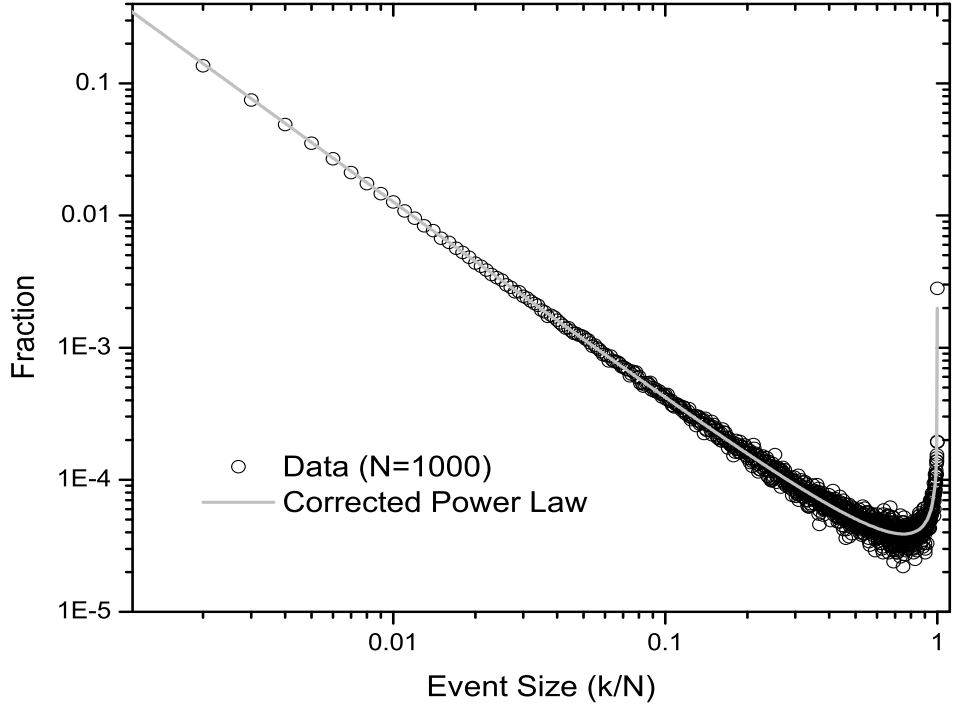}%
\caption{Theoretical event size distribution for a no-weakening critical model
with the additional constraint on the stochastic component of the stress
distribution. This results in an exact fit to the simulation data.}%
\label{Fig: Corrected Power Law}%
\end{center}
\end{figure}

\begin{figure}
[ptbptb]
\begin{center}
\includegraphics[
height=4.5637in,
width=5.9153in
]%
{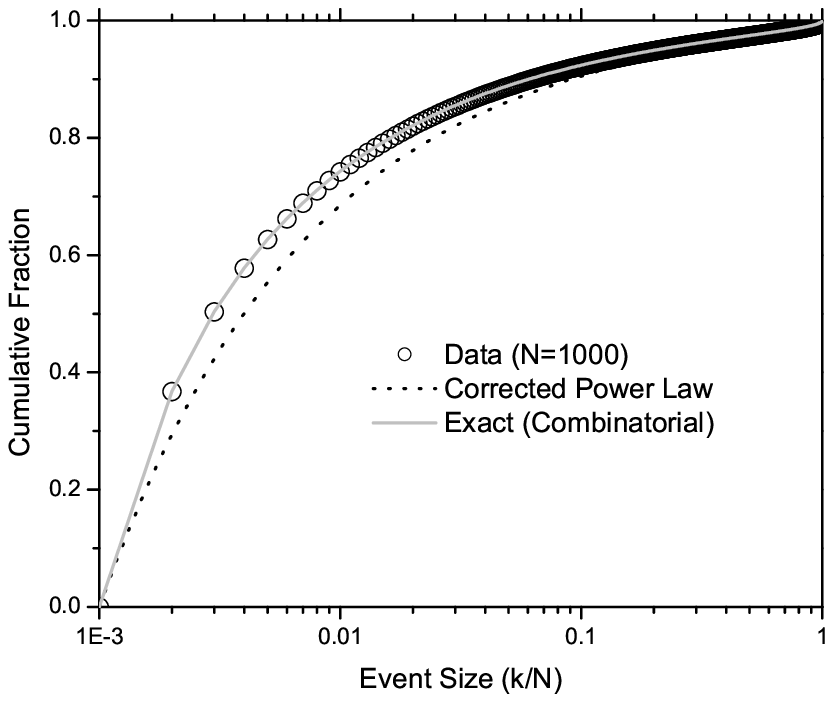}%
\caption{The Cumulative distribution of events compared to the theoretical
power law. Discrepancies for small events show the limits of Stirling's
approximation. Using the combinatorial expression yields an exact fit to
simulation data.}%
\label{Fig: Cumulative Power Law}%
\end{center}
\end{figure}

To correct this we weight each probability $P_{0}(k)$ by the probability that
a walk starting at that point will return to zero at $k=N.$ This latter
quantity is $P_{0}(N-k)$. Dividing the product of these terms by $P_{0}(N)$
provides the proper normalization. The resulting constrained distribution
$P_{0}^{\prime}(k)$ is given by
\begin{equation}
P_{0}^{\prime}(k)=\frac{(2k)!(2N-2k)!N!^{2}}{(2N)!(N-k)!^{2}k!^{2}%
}.\label{exact constrained distribution}%
\end{equation}
The approximated first passage distribution $P_{0}^{\prime}(k)$ is shown in
Fig. \ref{Fig: Corrected Power Law}. This is well approximated by a corrected
power law%
\begin{equation}
P_{0}^{\prime}(\kappa)=\frac{\kappa^{-3/2}}{\sqrt{2\pi}\sqrt{1-\kappa}%
}.\label{Corrected Power Law}%
\end{equation}
\ Note this has a minimum at exactly $\kappa=3/4$. We conclude that this
feature, which resembles a nucleation phenomenon, stems from the basic
statistics of the stress distribution. For models with a large number of
sites, the low number of large events often makes it difficult to compare the
data to an exact power law. However, we can also compare the cumulative stress
distribution the the integral of (\ref{Corrected Power Law}). This is shown in
Fig. \ref{Fig: Cumulative Power Law}. However, a discrepancy at small event
sizes, due to error in Stirling's approximation is evident in this graph.
Using the exact combinatorial expression (\ref{exact constrained distribution}%
) instead yields a perfect fit to the simulation data.

Note in this particular case we can also derive the distribution from first
principles, since each site contributes exactly $1/N$ to the stress transfer,
when the rupture arrests there must be exactly $k$ sites in an interval $k/N$,
which for $N$ uniformly distributed variables can be written%
\begin{equation}
P^{\prime}(k)=\frac{1}{k}\left(
\begin{array}
[c]{c}%
N\\
k
\end{array}
\right)  \left(  \frac{k}{N}\right)  ^{k}\left(  1-\frac{k}{N}\right)  ^{N-k}.
\end{equation}
However, the above treatment is more general, and may be used to consider the
distribution of event sizes for non-critical models.

When $K_{R}>\varepsilon$, The idealized stress excess is negative (and
diverges at $\kappa=K_{R}+1$). The magnitude of the stochastic term must
overcome this deficit for a rupture to propagate. A previous approach
\cite{Dahmen98} is to linearize the stress excess about $\kappa=0$ (not
including terms of $\mathcal{O}(\varepsilon)$)%

\begin{equation}
v(\kappa)\simeq-\frac{K_{R}}{1+K_{R}}\kappa-\frac{X_{\kappa}}{\sqrt{N}}%
\end{equation}
and compute the first crossings of the random walk with the line. The exact
result is given by the Bachelier-Levy formula \cite{Lerche86}. We can derive
the result by defining a biased walk where an step up has probability
$p=\frac{1}{2}-\frac{c}{\sqrt{N}}$ and a step down $q=\frac{1}{2}+\frac
{c}{\sqrt{N}}$, where $c=\frac{K_{R}}{2(K_{R}+1)}$. The first crossing
probabilities for $k=1,2,\ldots$ are
\begin{equation}
\frac{P_{0}(k)}{k}=\frac{1}{k}\frac{(2k)!}{k!^{2}}\left(  \frac{1}{4}%
-\frac{c^{2}}{N}\right)  ^{k}\thicksim k^{-3/2}e^{-4c^{2}\kappa},\;N>>1.
\end{equation}
For the limiting value of $c=1/2$ ($K_{R}\rightarrow\infty$), the power law is
dominated by an exponential decay $e^{-\kappa}$.

However, the linearized stress excess does not provide a valid approximate
solution. The actual distribution will be truncated much more quickly.
Ignoring terms of $\mathcal{O}(\varepsilon)$ we can write the entire
expression for the stress excess%
\begin{align}
\nu(\kappa)  &  \simeq-\frac{K_{R}}{K_{R}+1}\kappa-\frac{K_{R}}{2(K_{R}%
+1)(K_{R}+1-\kappa)}\kappa^{2}\nonumber\\
&  =-\frac{K_{R}}{2}\left[  \frac{\kappa}{K_{R}+1}+\frac{\kappa}%
{K_{R}+1-\kappa}\right]
\end{align}
Note that the magnitude of the nonlinear term overtakes the first when
$\kappa>(K_{R}+1)/2$. The second line gives the stress excess function in its
most compact form.

\begin{figure}
[ptb]
\begin{center}
\includegraphics[
height=4.5637in,
width=5.9153in
]%
{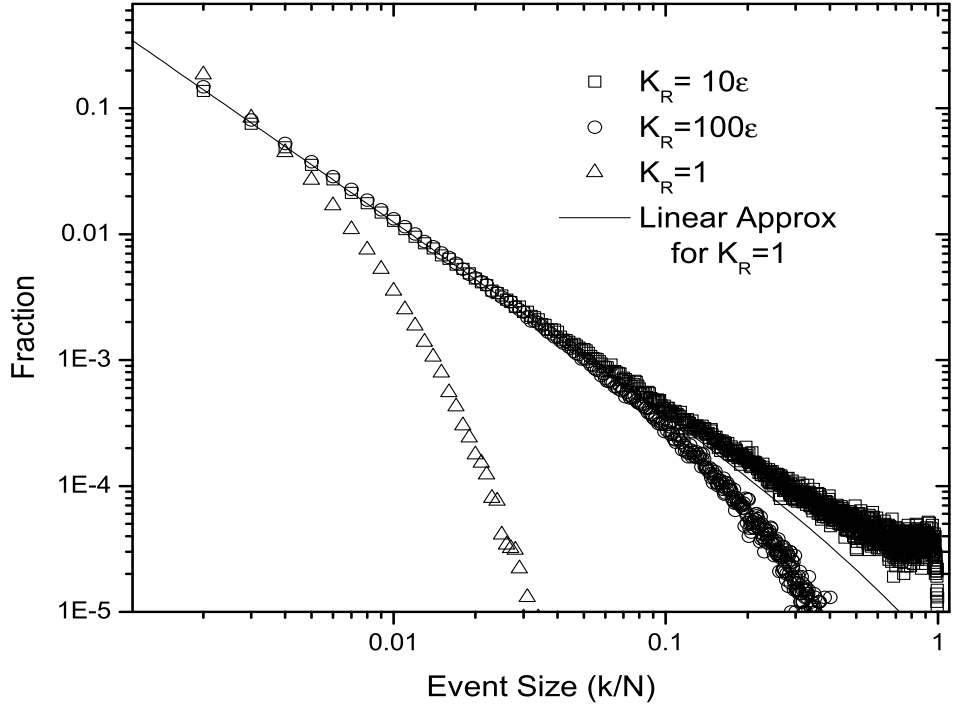}%
\caption{Event size distributions for simulations with $K_{R}>\varepsilon$.
Also shown is the theoretical distribution based on the linearized stress
excess for $K_{R}=1$. The data falls off much more rapidly due to the
nonlinear term.}%
\label{Fig: Truncated}%
\end{center}
\end{figure}

If we label the non-stochastic part of the stress excess $\nu(\kappa
)=\nu^{\prime}(\kappa)+X_{\kappa}/\sqrt{N}$, the event size distribution can
be formulated as the distribution of first intersection (Fig.
\ref{Fig: Truncated}).%
\begin{equation}
X_{\kappa}=\sqrt{N}\,\left\vert \nu^{\prime}(\kappa)\right\vert
\end{equation}
where we recognize the sign of the walk term is irrelevant. There is an
extensive literature regarding the intersection of a random walk with a
curvilinear boundary \cite{Lerche86,Durbin85}. Several formal solutions are
available, but analytic forms exist only for special cases. Numeric
computation based on the formal solution is still useful because it eliminates
sampling errors for low probability events.

\section{Conclusions}

The forced weakening method introduces a means of modeling dynamical behavior
leading to an instability in an efficient discrete time simulation. This is
done by using an arbitrary weakening function to compute stress transfers
during simulation, then assigning random residual stresses consistent with the
weakening function at the completion of rupture. This technique results in
Abelian avalanches allowing a rigorous and implementation-independent
analysis. Event distributions derived from this analysis yield excellent
agreement between theory and simulation, including a novel explanation for the
overabundance of large events. If one used a small-scale dynamical model to
characterize the average internal stress of a rupture as it depends on the
physical parameters of a rate and state dependent frictional law, we could
understand features of the resulting event size distribution.

This approach should be equally applicable to related discrete threshold
models. While the resulting formalism is dependent on the mean-field character
of the model, the numerical techniques may find wider use whenever an
inversion like (\ref{Stress Transfer 2}) is available. Use of a stress excess
function to understand the event size distribution may be extendible to non
mean field models with a finite interaction range. Local correlations might
result in the stochastic component resembling a fractional Brownian motion,
predictably altering the crossing distribution.

E.F.P. and J.B.R. acknowledge support by DOE grant DE-FG03-95ER14499; J.S.S.M.
was supported as a Visiting Fellow by CIRES, University of Colorado at Boulder.

\end{document}